\documentstyle[12pt]{article}
\def\rf#1{(\ref{eq:#1})}
\def\lab#1{\label{eq:#1}}
\def\nn{\nonumber \\}
\newcommand{\beano}{\begin{eqnarray*}}
\newcommand{\enano}{\end{eqnarray*}}
\def\bea{\begin{eqnarray}}
\def\ena{\end{eqnarray}}
\def\be{\begin{equation}}
\def\ee{\end{equation}}
\def\foot#1{\footnotemark\footnotetext{#1}}

\def\pv{\vec{p}}
\def\kv{\vec{k}}
\relax
\def\ra{\rightarrow}

\def\NPB#1#2#3{{\sl Nucl. Phys.} {\bf B#1} (#2) #3}

\def\PRv#1#2#3{{\sl Phys. Rev.} {\bf #1} (#2) #3}

\def\JMP#1#2#3{{\sl J. Math. Phys.} {\bf #1} (#2) #3}

\def\AoP#1#2#3{{\sl Ann. of Phys.} {\bf #1} (#2) #3}

\def\IJMPA#1#2#3{{\sl Int. J. Mod. Phys.} {\bf A#1} (#2) #3}

\def\NC#1#2#3{{\sl Nuovo Cimento} {\bf #1} (#2) #3}
\def\NP#1#2#3{{\sl Nucl. Phys.} {\bf #1} (#2) #3}
\begin{document}
\begin{titlepage}
\vspace{-1cm}
\noindent
\hfill{US-FT/3-99}\\
\vspace{0.0cm}
\hfill{hep-th/9901145}\\
\vspace{0.0cm}
\hfill{March 1999}\\
\vspace{-0.4cm}
\hfill{(Revised version)}\\
\phantom{bla}
\vfill
\begin{center}
{\large\bf  Hermitian analyticity versus Real analyticity}\\
\vspace{0.3cm}
{\large\bf  in two-dimensional factorised $S$-matrix theories}
\end{center}

\vspace{0.3cm}
\begin{center}
J. Luis Miramontes
\par \vskip .1in \noindent
{\em 
Departamento de F\'\i sica de Part\'\i culas,\\
Facultad de F\'\i sica\\
Universidad de Santiago de Compostela\\
E-15706 Santiago de Compostela, Spain}\\
\par \vskip .1in \noindent
e-mail: miramont@fpaxp1.usc.es
\normalsize
\end{center}
\vspace{.2in}
\begin{abstract}
\vspace{.3 cm}
\small
\par \vskip .1in \noindent
The constraints implied by analyticity in two-dimensional factorised
$S$-matrix theories are reviewed. Whenever the theory is not
time-reversal invariant, it is argued that the familiar
condition of `Real analyticity' for the  $S$-matrix amplitudes has to be
superseded by a different one known as `Hermitian analyticity'. Examples
are provided of integrable quantum field theories whose
(diagonal) two-particle $S$-matrix amplitudes are Hermitian analytic but
not Real analytic. It is also shown that Hermitian analyticity is
consistent with the bootstrap equations and that it ensures the
equivalence between the notion of unitarity in the quantum group approach
to factorised $S$-matrices and the genuine unitarity of the $S$-matrix.

\end{abstract}
\vfill
\end{titlepage}
It is well known that the $S$-matrix of an integrable two-dimensional
quantum field theory factorises into products of two-particle
amplitudes. Then, the property of factorisation itself and the usual
`axioms' of $S$-matrix theory constrain the allowed form of the
$S$-matrix to such an extent that it becomes possible to
conjecture its form. Namely, consistency with factorisation
translates into the `Yang Baxter equation' while, on general grounds, it
is assumed that the $S$-matrix exhibits `unitarity', `crossing
symmetry', `analyticity', and satisfies the `bootstrap
equations'~\cite{SMAT} (for a nice recent review see~\cite{SMATpat} and
the references therein).

In addition to this, and especially when the $S$-matrix is
diagonal, many works on factorised $S$-matrices assume `Real
analyticity'. This means that two-particle amplitudes are not only
analytic functions of their arguments, but they take complex-conjugate
values at complex-conjugate points. However, this is not required by the
usual $S$-matrix axioms.

The aim of this letter is to review the consequences of
analyticity for two-dimensional $S$-matrix amplitudes, and to serve as
a reminder that Real analyticity is not essential. In the framework of
$S$-matrix theory, it is  just a special case of a general
condition known as `Hermitian analyticity'~\cite{OLHA} which is valid
only when the theory happens to be time-reversal symmetric. For the
two-particle $S$-matrix amplitudes, we will deduce the form of the
Hermitian analyticity constraints which are given by eq.~\rf{HASgen} and
constitute our main result.

We provide several examples of diagonal $S$-matrices which are Hermitian
analytic but not Real analytic. All of them correspond to non-parity
invariant integrable field theories. The simplest is a fermion model
proposed by Federbush in 1960~\cite{FED,FEDb}. The others are different
quantum field theories associated with the non-abelian affine Toda
equations recently constructed in~\cite{MASS,PARA,BRA}. In all these
cases, the scattering amplitudes between different particles are not Real
analytic. Instead, those amplitudes connected through a parity
transformation are related in such a way that Real analyticity would be
satisfied if the theory becomes parity invariant. This unusual
analyticity condition is just a consequence of Hermitian analyticity.

Other cases where Real analyticity is not satisfied
arise in the quantum group approach to factorised $S$-matrices~\cite{QGS}
and, in particular, to (abelian) affine Toda field theories with imaginary
coupling constant~\cite{MATRIX,QGToda,GEOR}. There, the $S$-matrix is a
quantum group $R$-matrix times a Real analytic scalar factor, which
implies that those amplitudes whose matrix structure is trivial are
Real analytic. In this approach, the relationship between the analytic
continuations of the different amplitudes is dictated by the properties
of the $R$-matrix and, in general, Hermitian analyticity is
not satisfied either. 

At this point it is important to mention that, in
the framework of standard $S$-matrix theory where Hermitian analyticity is
formulated, the axiom of `unitarity' actually means that the $S$-matrix
is unitary: $S^\dagger S=1$. In contrast, in the quantum group approach
the role of unitarity is played by another condition called `$R$-matrix
unitarity' (RU). In fact, Tak\'acs and Watts have recently highlighted
that some of the resulting $S$-matrices are not unitary, which does not
prevent them describing physically relevant (non-unitary)
models~\cite{TW}. We will show that Hermitian analyticity ensures that RU
is equivalent to physical unitarity without any extra requirements. In
other words, Hermitian analyticity is a sufficient condition to guarantee
that a factorised $S$-matrix obtained through the quantum group
approach is unitary.

An excellent classical review of analyticity in $S$-matrix theory is
provided by the book of Eden {\em et al.\/}~\cite{EDEN}, which will be
our main reference in the following. Analyticity is the assumption that
the physical $S$-matrix amplitudes are real boundary values of
analytic functions as a consequence of causality and the existence of
macroscopic time. On top of this, the unitarity equations are expected 
to evaluate the discontinuities of those analytic functions across their
normal-threshold cuts. This requires that the physical $S$- and
$S^\dagger$-matrix amplitudes are opposite boundary values of the {\em
same} analytic functions, which states the property known as
Hermitian analyticity~\cite{OLHA,EDEN}.

Let us consider a generic integrable theory whose spectrum consists of
several degenerate multiplets labelled by a set of finite dimensional
vector spaces $V_A, V_B, \ldots$ with different masses $m_A, m_B,
\ldots$. Particles in the same multiplet will be distinguished by a
flavour index `$i$' and, for simplicity, we will assume that all these
particles are bosonic. Since the theory is integrable, the only
non-vanishing (connected) two-particle $S$-matrix amplitudes are of the
form
\be
\langle \kv_{A_k}, \kv_{B_l}\mid S - 1\mid  \pv_{A_i},
\pv_{B_j}\rangle
\> = \> (2\pi)^2
\> \delta^{(2)}(p_{A_i}+ p_{B_j}- k_{A_k} - k_{B_l})\>
i\> {\cal M}_{ij\ra kl}^{AB}\>,
\lab{Mampl}
\ee
where $|\pv_{A_i}, \pv_{B_j}\rangle$ is the state of two
particles with mass $m_A$ and $m_B$, and momentum $\pv_{A_i}$ and
$\pv_{B_j}$:
\be   
|\pv_{A_i},  \pv_{B_j}\rangle\> =\> a_{A_i}^\dagger
(\pv_{A_i})  \> a_{B_j}^\dagger(\pv_{B_j}) |0\rangle\>.
\ee
Lorentz invariance allows one to decompose the scattering amplitude
into scalar and pseudoscalar parts:
\be 
{\cal M}_{ij\ra kl}^{AB}\> =\> M_{ij\ra kl}^{AB}(s) \> +\>4
\epsilon_{\mu \nu}\> p_{A_i}^{\mu}  p_{B_j}^{\nu}\> P_{ij\ra
kl}^{AB}(s)\>,
\ee
where $M_{ij\ra kl}^{AB}(s)$ and $P_{ij\ra kl}^{AB}(s)$ are functions
of the Mandelstam variable $s= (p_{A_i} + p_{B_j})^2$ only.

Analyticity postulates that the scalar and pseudoscalar
components of the scattering amplitudes, $M_{ij\ra kl}^{AB}$ and
$P_{ij\ra kl}^{AB}$, are boundary values of analytic functions. This
means that they can be continued to complex values of
$s$, and that the resulting functions are analytic. In this case,
since the theory is integrable, they should
exhibit only two cuts along $s\geq (m_A+ m_B)^2$ and
$s\leq (m_A- m_B)^2$ on the real axis with square root branching points,
corresponding to the physical processes in the $s$- and $t$-channel,
respectively. Then, the physical $s$-channel amplitudes are
given by the limit onto the cut from the upper-half complex $s$-plane,
\be
M_{ij\ra kl}^{{AB}^{\;\rm phys}}(s) \> =
\lim_{\epsilon\rightarrow 0^+}
\>  M_{ij\ra kl}^{AB}(s+ i\epsilon)\>, \qquad
P_{ij\ra kl}^{{AB}^{\;\rm phys}}(s) \> =
\lim_{\epsilon\rightarrow 0^+}
\>  P_{ij\ra kl}^{AB}(s+ i\epsilon)\>,
\ee 
which is the generalization of the well known Feynman's $i\epsilon$
prescription in perturbation theory.

Hermitian analyticity goes one step beyond. It postulates that the
physical $S$- and $S^\dagger$-matrix amplitudes are opposite boundary
values of the {\em same} analytic functions, a property
that has been proved in perturbation theory~\cite{OLHA}, in
potential theory~\cite{TAYLOR}, and using $S$-matrix theory
alone~\cite{OLHAb} (see~\cite{EDEN} and the references therein).
Since
\be
\langle \kv_{A_k}, \kv_{B_l}\mid S^\dagger - 1\mid 
\pv_{A_i}, \pv_{B_j}\rangle\> =
\langle \pv_{A_i}, \pv_{B_j} \mid S - 1\mid  \kv_{A_k},
\kv_{B_l}\rangle^\ast
\>,
\ee
this condition can be written as
\be
\Bigl[M_{kl\ra ij}^{{AB}^{\;\rm phys}}(s)\Bigr]^\ast \> =
\lim_{\epsilon\rightarrow 0^+}
\>  M_{ij\ra kl}^{AB}(s- i\epsilon)\>,
\ee
and a similar equation for $P_{ij\ra kl}^{AB}$. 
Therefore, taking into account that both $M_{ij\ra kl}^{AB}(s)$ and
$M_{kl\ra ij}^{AB}(s)$ are analytic functions  of $s$, and
using that if $f(z)$ is analytic so also is
$g(z)=[f(z^\ast)]^\ast$, Hermitian analyticity results in the
following relationships:
\be
M_{ij\ra kl}^{AB}(s) \> = \> \bigl[M_{kl\ra
ij}^{AB}(s^\ast)\bigr]^\ast \>, \qquad
P_{ij\ra kl}^{AB}(s) \> = \> \bigl[P_{kl\ra
ij}^{AB}(s^\ast)\bigr]^\ast \>.
\lab{HA}
\ee
An immediate and vital consequence of Hermitian analyticity is
that the unitarity equations $S^\dagger S=1$ evaluate the 
discontinuities of $M_{ij\ra kl}^{AB}(s)$ and $P_{ij\ra kl}^{AB}(s)$
across the two-particle cuts~\cite{OLHA,STAPP}.

In two-dimensions, it is customary to use rapidities instead of
momenta, 
\be
(p_{A_i}^0, \pv_{A_i})\> =\> (m_A \cosh \theta_{A_i}\>, \> m_A \sinh
\theta_{A_i}) \>.
\ee
Then, the (real) Mandelstam variable~$s$ is a function of the absolute
value of the rapidity difference of the colliding particles $\theta =
|\theta_{A_i} -\theta_{B_j}|>0$, 
\be 
s\> =\> (p_a\>+ \>p_b)^2 \> =\> m_a^2 \>+ \> m_b^2\> +\> 2\>
m_a\> m_b\> \cosh\theta\>,
\ee
and the two-particle amplitudes become functions of $\theta$.
Understood as complex variables, the change of variables between $s$
and $\theta$ allows one to open the two cuts. Hence, $M_{ij\ra
kl}^{AB}(\theta)$ and $P_{ij\ra kl}^{AB}(\theta)$ are meromorphic and 
the physical sheet is mapped into the region
$0\leq {\rm Im\/} \theta \leq \pi$, which is the first Riemann sheet 
in the complex $\theta$-plane. 

Regarding analyticity, notice that
\be
\lim_{\epsilon\rightarrow 0^+} \> 
M_{ij\ra kl}^{AB} (s\pm i\epsilon)\> = \>M_{ij\ra
kl}^{AB}(\pm \theta)\>, \quad \theta>0\>.
\ee
Therefore, since the amplitudes are meromorphic functions of $\theta$,
the Hermitian analyticity relationships~\rf{HA} translate into
\be
M_{ij\ra kl}^{AB}(\theta)\> = \> \bigl[M_{kl\ra
ij}^{AB}(-\theta^\ast)\bigr]^\ast\>, \qquad
P_{ij\ra kl}^{AB}(\theta)\> = \> \bigl[P_{kl\ra
ij}^{AB}(-\theta^\ast)\bigr]^\ast\>.
\lab{HAtheta}
\ee

Finally, let us consider the full $S$-matrix amplitude
corresponding to~\rf{Mampl}:
\be
\langle \kv_{A_k}, \kv_{B_l}\mid S \mid  \pv_{A_i},
\pv_{B_j}\rangle =  4 (2\pi)^2  \delta(\theta_{A_i}
-\theta_{A_k})
\delta(\theta_{B_j} -\theta_{B_l}) \> {\cal S}_{ij
\ra kl}^{AB}\>,
\ee
where
\bea
&& {\cal S}_{ij \ra kl}^{AB}\> =\> \delta_{ik} \delta_{jl}+
i {{\cal M}_{ij \ra kl}^{AB} \over 4 m_A m_B \sinh \theta}  \nn
\noalign{\vskip 0.4truecm}
& & \qquad = \delta_{ik} \delta_{jl}+
 i\left( {M_{ij\ra kl}^{AB}(\theta) \over 4 m_A m_B \sinh \theta}\> + \>
P_{ij\ra kl}^{AB}(\theta)\> {\rm sign\/}(\theta_{A_i}
-\theta_{B_j})\right)
\>, 
\lab{SmatBig}
\ena
and we have used that
\be
4\epsilon_{\mu \nu}\> p_{A_i}^{\mu}  p_{B_j}^{\nu}\> = \>
4m_A m_B \sinh (\theta_{A_i} -\theta_{B_j})\> =\> 
4m_A m_B \sinh \theta \> {\rm sign\/}(\theta_{A_i}
-\theta_{B_j})\>,
\ee
together with the standard relativistic normalization
\be
\langle \pv_{B_j}\mid \pv_{A_i}\rangle\> = \> \delta_{AB}
\>\delta_{ij}\> (2\pi)
\> 2 p_{A_i}^0\> \delta(\pv_{A_i}- \pv_{B_j})\>.
\ee

Using the Heaviside function $\vartheta(x) = 0$ if $x<0$ and $=1$ if
$x>0$, eq.~\rf{SmatBig} can be written as 
\be
{\cal S}_{ij \ra kl}^{AB} \> =\> \vartheta(\theta_{A_i} -\theta_{B_j})\> 
{S_{AB}\>}_{ij}^{kl}(\theta) \> + \>
\vartheta (\theta_{B_j}-\theta_{A_i}) \> {S_{BA}\>}_{ji}^{nm}(\theta)\> ,
\ee
where
\be 
{S_{AB}\>}_{ij}^{kl}(\theta)\> =\> \delta_{ik} \delta_{jl}\> +\> 
 i\left( {M_{ij\ra kl}^{AB}(\theta) \over 4 m_A m_B \sinh \theta}\> +\>
P_{ij\ra kl}^{AB}(\theta) \right)\>, 
\lab{DefL}
\ee
is the scattering amplitude of the process where particle
$A_i$ initially is on the left-hand side of particle $B_j$, while
\be 
{S_{BA}\>}_{ji}^{lk}(\theta)\> =\> \delta_{ik} \delta_{jl} \> +\>
 i\left( {M_{ij\ra kl}^{AB}(\theta) \over 4 m_A m_B \sinh \theta}\> -
\> P_{ij\ra kl}^{AB}(\theta)\right)\>, 
\lab{DefR}
\ee
is the amplitude of the process where $A_i$ initially is on the
right-hand side of $B_j$. These amplitudes can be seen as the
matrix elements of two maps~\cite{MATRIX}
\be
S_{AB}(\theta) :\> V_A \otimes V_B \longrightarrow V_B \otimes V_A\> ,
\qquad S_{BA}(\theta) :\> V_B \otimes V_A \longrightarrow V_A \otimes
V_B\>,
\lab{Matrix}
\ee
where $\theta$ is the rapidity difference of the incoming particles.
Equivalently, in the symbolic algebraic notation commonly used to describe
two-dimensional factorised $S$-matrix theories~\cite{SMAT,SMATpat}, they
correspond to
\bea
&& A_i(\theta)\> B_j(\theta')\> =\> \sum_{k,l}
{S_{AB}\>}_{ij}^{kl} (\theta-\theta') \> B_l(\theta') \>
A_k(\theta)\>, \nn 
\noalign{\vskip 0.1truecm}
&& B_j(\theta)\> A_i(\theta')\>  =\> \sum_{k,l}
{S_{BA}\>}_{ji}^{lk} (\theta-\theta') \> A_k(\theta')\>
B_l(\theta) \>.
\lab{Symbol}
\ena

The two-particle amplitudes ${S_{AB}\>}_{ij}^{kl}$ and 
${S_{BA}\>}_{ji}^{lk}$ are analytic functions of $\theta$. Moreover,
taking into account~\rf{HAtheta}, \rf{DefL}, and~\rf{DefR}, they satisfy
\be
{S_{AB}\;}_{ij}^{kl} (\theta)\> = \> \Bigl[{S_{BA}\;}_{lk}^{ji}
(-\theta^\ast)\Bigr]^\ast
\lab{HASgen}
\ee
which summarises Hermitian analyticity in two-dimensional factorised
$S$-matrix theories and is our central result. Eq.~\rf{HASgen} means
that the two maps defined in~\rf{Matrix} are related according to
$ S_{AB}(\theta ) \> =\> S_{BA}^\dagger (-\theta^\ast)$, 
where the dagger stands for Hermitian conjugation.

A direct consequence of~\rf{HASgen} is that the scattering
amplitudes will not be Real analytic functions unless they exhibit
additional symmetry properties. To spell this out, recall the behaviour
of the two-particle $S$-matrix amplitudes with respect to parity (P) and
time-reversal (T) transformations:
\be
{\rm P}:\> {S_{AB}\;}_{ij}^{kl} (\theta) \longrightarrow
{S_{BA}\;}_{ji}^{lk} (\theta) \>, \qquad
{\rm T}:\> {S_{AB}\;}_{ij}^{kl} (\theta) \longrightarrow
{S_{BA}\;}_{lk}^{ji} (\theta) \>.
\ee
This shows that Real analyticity is a special case of Hermitian
analyticity which is valid only when the amplitude happens to be
symmetric with respect to time-reversal transformations, a
conclusion that could have been anticipated on general
grounds~\cite{EDEN}. 

Another important consequence of~\rf{HASgen} concerns the
formulation  of the unitarity condition. For real $\theta>0$, the
unitarity of the $S$-matrix, $S^\dagger S= 1$, translates into
\be
\sum_{k,l} {S_{AB}\;}_{ij}^{kl} (\theta) \>
\Bigl[{S_{AB}\;}_{i'j'}^{kl} (\theta)\Bigr]^\ast \> =\>
\delta_{ii'} \> \delta_{jj'}\>.
\ee
However, using the Hermitian analyticity condition~\rf{HASgen},
unitarity can be equivalently written as
\be
\sum_{k,l} {S_{AB}\;}_{ij}^{kl} (\theta) \>
{S_{BA}\;}_{lk}^{j'i'} (-\theta) \> =\>
\delta_{ii'} \> \delta_{jj'}\>,
\lab{RU}
\ee
which is nothing else than the condition of `$R$-matrix unitarity'
(RU) that arises naturally in the quantum group approach to
factorised $S$-matrices~\cite{QGS,MATRIX,QGToda,GEOR,TW}. Actually, to be
precise, RU is the analytic continuation of~\rf{RU} to the complex
$\theta$-plane. Therefore, we conclude that there is no
difference between physical unitarity and the quantum group inspired
$R$-matrix unitarity if the $S$-matrix amplitudes exhibit Hermitian
analyticity.

In order to validate the Hermitian analyticity condition
given by eq.~\rf{HASgen}, it is necessary to check whether it is preserved
by the bootstrap equations. Suppose that
${S_{AB}\>}_{ij}^{kl}(\theta)$ has a simple pole at $\theta= iu_{AB}^C$
on the physical strip corresponding to a bound state in the multiplet
$V_{\overline{C}}$. Thus, its residue is provided by the projector of
$V_A\otimes V_B$ into $V_{\overline{C}}\subset V_B\otimes V_A$ and, near
the pole, the amplitude will be of the form
\be
{S_{AB}\>}_{ij}^{kl}(\theta)\>  \sim \>{i
\over \theta\> -\>iu_{AB}^C}\> \sum_a G_{ij}^{a}\> {H_{l
k}^{a}}^\ast\>,
\lab{PoleAB}
\ee
where, in the symbolic notation of~\cite{SMAT,SMATpat} already used in
eq.~\rf{Symbol}, the coupling constants $G_{ij}^{a}$ and $H_{ji}^{a}$
are defined through the identities~\foot{In these equations, we use the
standard notation for the fusion angles such that $\overline{u}_{AB}^C =
\pi - u_{AB}^C$, $\overline{u}_{AC}^B +
\overline{u}_{BC}^A +\overline{u}_{AB}^C =\pi$, and $m_{\overline{C}} =
{\rm e\>}^{+i\overline{u}_{AC}^B} m_A + {\rm e\>}^{-i\overline{u}_{BC}^A}
m_B$.}
\bea
&&
\lim_{\theta_1-\theta_2\ra iu_{AB}^C }\> \bigl(\theta_1 -\theta_2
-iu_{AB}^C\bigr)\> A_i(\theta_1)\> B_j(\theta_2)\> =\>
i\> \sum_a G_{ij}^{a}\>\> \overline{C}_a\Bigl({\theta_1 -
i \overline{u}_{AC}^B + \theta_2 +i\overline{u}_{BC}^A  \over2}\Bigr)\>,
\nn
\noalign{\vskip 0.2truecm}
&&
\qquad\qquad\qquad
\overline{C}_a(\theta)\> =\> \sum_{kl}
{H_{lk}^{a}}^\ast\>  B_l(\theta -i\overline{u}_{BC}^A)\> A_k(\theta
+i\overline{u}_{AC}^B)\>.
\lab{ProA}
\ena 
Correspondingly, using the Hermitian analyticity condition, 
\be
{S_{BA}\>}_{lk}^{ji}(\theta)\> =\>
\bigl[{S_{AB}\>}_{ij}^{kl}(-\theta^\ast)\bigr]^\ast \sim \>{i
\over \theta\> -\>iu_{AB}^C}\> \sum_a H_{lk}^{a} \>{G_{ij}^{a}}^\ast \>,
\lab{PoleBA}
\ee
which shows that the amplitude ${S_{BA}\>}_{lk}^{ji}(\theta)$ also
exhibits a simple pole at the same location. Moreover, it provides
an equivalent definition of the coupling constants through the projection
of $V_B\otimes V_A$ into $V_{\overline{C}}\subset V_A\otimes
V_B$:
\bea
&&
\lim_{\theta_1-\theta_2\ra iu_{AB}^C }\> \bigl(\theta_1 -\theta_2 
-iu_{AB}^C\bigr)\> B_l(\theta_1)\> A_k(\theta_2)\> =\>
i\>
\sum_a H_{lk}^{a}\> \> \overline{C}_a\Bigl({\theta_1 -
i \overline{u}_{BC}^A + \theta_2 +i\overline{u}_{AC}^B  \over2}\Bigr)\>,
\nn
\noalign{\vskip 0.2truecm}
&& \qquad\qquad\qquad
\overline{C}_a(\theta)\> =\> \sum_{ij} {G_{ij}^{a}}^\ast\> 
A_i(\theta -i\overline{u}_{AC}^B)\> B_j(\theta +i\overline{u}_{BC}^A)\>,
\lab{ProB}
\ena 
In other words, $G_{ij}^{a}$ and $H_{ji}^{a}$ are the coupling constants
of the fusions $A_i B_j\ra {\overline C}_a$ and $B_j A_i \ra {\overline
C}_a$,  respectively, and, since the amplitudes are not always parity
symmetric, $G_{ij}^{a}\not=H_{ji}^{a}$ in general. Eqs.~\rf{PoleAB}
and~\rf{PoleBA} imply that, near the pole, the map $S_{AB}(\theta)$ is
of the form
\be
S_{AB}(\theta)\>  \sim \>{i
\over \theta\> -\>iu_{AB}^C}\> {P_{BA}^{C}}^\dagger\> P_{AB}^{C} \>,
\lab{Pole}
\ee
where $P_{AB}^{C} : V_A\otimes V_B \longrightarrow
V_{\overline{C}}$ is a projection operator. Eq.~\rf{Pole} is explicitly
consistent with the Hermitian analyticity condition and
manifests that the poles in $S_{AB}(\theta)$ and $S_{BA}(\theta)$
correspond to particles in the same multiplet $V_{\overline{C}}$.

The bootstrap equations express the fact that there
is no difference whether the scattering process with any particle in, say,
$V_D$ occurs before or after the fusion of particles $A_i$ and $B_j$ into
particle ${\overline C}_a$. In our case, using eqs.~\rf{ProA}
and~\rf{ProB}, this condition allows one to write four different but
equivalent expressions for the scattering amplitudes involving the
particles in $V_{\overline{C}}$ and $V_D$. For our purposes, it will be
enough to consider only the following two 
\bea
\sum_b {S_{\overline{C}D}\>}_{am}^{bn}(\theta)  \> {H_{lk}^{b}}^\ast &=&
\sum_{ij} {H_{ji}^{a}}^\ast \> {S_{AD}\>}_{im}^{kp}(\theta\>+ \>
i\overline{u}_{AC}^B)\>  {S_{BD}\>}_{jp}^{ln}(\theta\>-
\>i\overline{u}_{BC}^A)\>, \nn
\sum_b H_{lk}^{b} \>{S_{D\overline{C}}\>}_{nb}^{ma}(\theta)  &=&
\sum_{ij} {S_{DB}\>}_{nl}^{pj}(\theta\>- \>
i\overline{u}_{BC}^A)\>  {S_{DA}\>}_{pk}^{mi}(\theta\>+
\>i\overline{u}_{AC}^B) \> H_{ji}^{a}\>. 
\ena
Then, if the scattering amplitudes for the particles in $V_A$,
$V_B$, and $V_D$ satisfy the condition~\rf{HASgen}, it is straightforward
to check that
\be
\sum_b {S_{\overline{C}D}\>}_{am}^{bn}(\theta)  \>
{H_{lk}^{b}}^\ast \>  =\> \sum_b \Bigl[H_{lk}^{b} \>
{S_{D\overline{C}}\>}_{nb}^{ma}(-\theta^\ast)\Bigr]^\ast  
\>,
\ee
which proves that the amplitudes ${S_{\overline{C}D}\>}_{am}^{bn}$ and
${S_{D\overline{C}}\>}_{nb}^{ma}$ obtained by means of the bootstrap
principle are also Hermitian analytic.

It is worth noticing that the Hermitian analyticity
constraints given by eq.~\rf{HASgen} cannot be satisfied in all possible
bases on $V_A, V_B, \ldots$. Consider the following
change of basis on $V_A$ and $V_B$: $A_i(\theta) \ra
\widetilde{A}_i(\theta) = \sum_p {L_A\>}_i^p (\theta) A_p(\theta)$ and
$B_j(\phi) \ra \widetilde{B}_j(\phi) = \sum_q {L_B\>}_i^q (\phi)
B_q(\phi)$, where $L_A(\theta)$ and $L_B(\phi)$ are invertible but
not necessarily unitary matrices. In the new basis, the  scattering
amplitude ${S_{AB}\;}_{ij}^{kl} (\theta-\phi)$ becomes
\be
{S_{\widetilde{A} \widetilde{B}}\;}_{pq}^{rs} (\theta-\phi)\>
=\>\sum_{i,j,k,l} {L_{A}\>}_{p}^i(\theta)\> {L_{B}\>}_{q}^j(\phi)\>
{S_{AB}\;}_{ij}^{kl} (\theta-\phi)\> {L_{B}^{-1}\>}_{l}^s(\phi)\>
{L_{A}^{-1}\>}_{k}^r(\theta)\>.
\ee
Then, if ${S_{\widetilde{A} \widetilde{B}}\;}_{ij}^{kl}$ satisfies the
Hermitian analyticity condition~\rf{HASgen} it is straightforward to
check that 
\be
\sum_{p,q} \> {M_{A}\>}_{i}^p(\theta)\> {M_{B}\>}_{j}^q(\phi)\>
{S_{A B}\;}_{pq}^{kl} (\theta-\phi)\>
 = \>
\sum_{r,s}\> \bigl[{S_{B A}\;}^{ji}_{sr}
(\phi^\ast-\theta^\ast)\bigr]^\ast \> {M_{A}\>}_{r}^k(\theta)\>
{M_{B}\>}_{s}^l(\phi)\>,
\lab{HAW}
\ee
where
\be
{M_A\>}^{i}_k (\theta)\> =\> \sum_p {L_A\>}^{i}_p(\theta) \>
\bigl[{L_A\>}^{k}_p(\theta^\ast)\bigr]^\ast
\quad {\rm and}\quad
{M_B\>}^{j}_l (\phi)\> =\> \sum_p {L_B\>}^{j}_p(\phi) \>
\bigl[{L_B\>}^{l}_p(\phi^\ast)\bigr]^\ast \>.
\lab{Split}
\ee
Therefore, given a set of two-particle scattering amplitudes, a
sufficient condition to ensure that there is a basis where they are
Hermitian analytic is that there exists two matrices $M_A$
and $M_B$ of the form given by eq.~\rf{Split} such that the
constraints~\rf{HAW} hold. Notice that, $M_A$
and $M_B$ are hermitian positive definite matrices for real values of
$\theta$. We will refer to this condition as `weak' Hermitian
analyticity, thus making reference to the fact that it generalizes a
condition found by Liguori, Mintchev and Rossi in the context of exchange
algebras~\cite{MINT}. There, the amplitudes ${S_{A
B}\;}_{ij}^{kl}(\theta)$ for real values of
$\theta$ provide the exchange factors, and weak Hermitian analyticity
arises as a sufficient condition to allow the construction of a unitary
scattering operator in a Fock representation of the algebra.

In the rest of the letter, we provide several examples of diagonal
$S$-matrix theories where Hermitian analyticity holds but Real analyticity
is not satisfied. In all these cases the multiplets are not degenerate
and, hence, no flavour indices are needed. Then, the Hermitian analyticity
condition~\rf{HASgen} simplifies to  
\be
S_{AB} (\theta)\> = \> \Bigl[S_{BA} (-\theta^\ast)\Bigr]^\ast\>.
\lab{HAS}
\ee
It is worth noticing that, in the diagonal case, the P and T
transformations of the $S$-matrix amplitudes are identical. This
explains why all our examples involve non-parity invariant theories. 

Our first example will be the Federbush
model~\cite{FED}, which was studied in great detail, among others, by
Schroer, Truong, and Weisz~\cite{FEDb}. The Federbush model describes 
two massive Dirac fields $\psi_I$ and $\psi_{II}$ whose interaction
Lagrangian is
\be
L_{\rm FM} \> =\> -\> 2\pi \lambda \> \epsilon_{\mu \nu}\>
J_{I}^\mu\> J_{II}^\nu \>,
\ee
where $J_{I}^\mu$ and $J_{II}^\nu$  are the conserved vector currents
of the Dirac fields. The two-particle $S$-matrix amplitudes of the
Federbush model are particularly simple and can be written
as~\cite{FEDb}
\be
S_{I,I}(\theta)\> = \> S_{II,II}(\theta)\> =\> 1 \>, \qquad
S_{I,II}(\theta)\> = \> {\rm e\>}^{-2\pi\lambda i}\>, \qquad
S_{II,I}(\theta)\> = \> {\rm e\>}^{+2\pi\lambda i}\>.
\ee
Since they are given just by rapidity independent
phase factors, the amplitudes $S_{I,II}$ and $S_{II,I}$ are clearly
not Real analytic. However, it is straightforward to check that they
satisfy eq.~\rf{HAS} or, in other words, that the $S$-matrix of the
Federbush model is Hermitian analytic.

The second example is provided by the integrable perturbation of the
$SO(3)_k$ Wess-Zumino-Witten model discussed by Brazhnikov
in~\cite{BRA}.~\foot{Using the construction of ref.~\cite{MASS},
the model of Brazhnikov can be described as a Symmetric Space
sine-Gordon (SSSG) theory associated with the compact type-I symmetric
space $SU(3)/SO(3)$, and it is expected that many other SSSG theories
exhibit similar properties.} The spectrum of stable fundamental (or
basic) particles consists just of two particles $\psi$ and $\vartheta$
associated with the two simple roots of $su(2)$, the Lie algebra of
$SO(3)$, whose mass can be taken to be different. The
two-particle scattering amplitude for $\psi$ and $\vartheta$ has been
calculated at tree-level in~\cite{BRA} and, properly normalized, the
result is
\be
S_{\psi\vartheta} (\theta) \> = \> 1\> + \> {2i\over k\sinh(\theta -
\theta_0)}\> +\> \cdots  \> =\> \bigl[S_{\vartheta\psi}
(-\theta^\ast)\bigr]^\ast \>,
\ee
where $\theta_0$ is a non-vanishing real constant whose value depends on
the coupling constants of the model. These tree-level
amplitudes are not Real analytic but, on the
contrary, they satisfy the Hermitian analyticity condition
eq.~\rf{HAS}.

As our last example, let us consider the Homogenous sine-Gordon (HSG)
theories constructed in~\cite{MASS}. There is a HSG theory for each
simple compact Lie group $G$ that corresponds to an integrable
perturbation of the conformal field theory (CFT) associated with
the coset $G_k/U(1)^{\times r_g}$, where $r_g$ is the rank of $G$,
or, equivalently, of the theory of level-$k$
$G$-parafermions~\cite{PARA}. The semiclassical
spectrum of stable particles of the HSG theories has been
obtained in~\cite{HSGSOL}. If the group $G$ is simply laced, the
spectrum consists of $k-1$ particles for each simple root
$\vec{\alpha}_i$ of $g$, the Lie algebra of $G$, whose masses are
given by
\be
M_{\vec{\alpha}_{i}} (n) \> =\> {k\over \pi}\> m_{\vec{\alpha}_{i}}
\> \sin\left(\pi\> n\over k\right)\>, \qquad i\> =\> 1,\ldots,
r_g\>, \qquad n\> =\> 1, \ldots, k-1\>.
\lab{HSGspec}
\ee
In this equation,
\be
m_{\vec{\alpha}_{i}}\> =\> 2m\> \sqrt{(\vec{\alpha}_{i} \cdot
\vec{\lambda}_+)\> (\vec{\alpha}_{i} \cdot
\vec{\lambda}_-)}\>,
\ee
the constant $m$ is the only dimensionful parameter
of the theory, and $\vec{\lambda}_\pm$ are continuous vector coupling
constants taking values in the fundamental Weyl chamber of the
Cartan subalgebra of~$g$. For a generic choice of
$\vec{\lambda}_\pm$, all these masses will be different, and
an exact diagonal $S$-matrix for these theories has been recently
proposed~\cite{NEW}. For our purposes, it will be enough to quote the
result for $G=SU(3)$. For the fundamental particles, corresponding to
$n=1$ in~\rf{HSGspec}, the two-particle amplitudes can be written as
\bea
& & S_{\vec{\alpha}_{j}, \vec{\alpha}_{j}} (\theta)\> =\> 
{\sinh {1\over2} \bigl(\theta \> +\>
{2\pi \over k} \> i\bigr) \over \sinh {1\over2} \bigl(\theta \> -\>
{2\pi \over k}\> i \bigr)}\>, \qquad j\> =\> 1,2\>, \nn
\noalign{\vskip 0.1truecm}
& &S_{\vec{\alpha}_{1}, \vec{\alpha}_{2}} (\theta)\> =\>   
{\rm e\/}^{\epsilon  {\pi \over k} i}\> 
{\sinh {1\over2} \bigl(\theta
\> -\sigma \> - \>{\pi \over k} \> i\bigr) \over \sinh {1\over2}
\bigl(\theta \> -\sigma \> +\>  { \pi \over k}\> i
\bigr)}\> = \bigl[S_{\vec{\alpha}_{2}, \vec{\alpha}_{1}}
(-\theta^\ast) \bigr]^\ast \>, 
\lab{HSGSM}
\ena
where the value of the real parameter $\sigma$ depends on the coupling
constants $\vec{\lambda}_\pm$, and $\epsilon$ can be taken to be $+1$ or
$-1$. Eq.~\rf{HSGSM} provides a set of diagonal two-particle $S$-matrix
amplitudes that satisfy unitarity, crossing symmetry, and the
bootstrap equations~\cite{NEW}. However, they are Hermitian
analytic and not Real analytic.

To sum up, our main point was to recall that Real analyticity is not
essential in $S$-matrix theory; it is just a special case of a general
property called Hermitian analyticity. Then, we have derived the
constraints implied by Hermitian analyticity for the two-particle
scattering amplitudes in two-dimensional factorised
$S$-matrix theories, which are summarised by eq.~\rf{HASgen}. 
These constraints are consistent with the bootstrap equations and agree
with the properties of the scattering amplitudes of several non-parity
invariant theories whose $S$-matrix is diagonal already
discussed in the literature~\cite{FED,FEDb,BRA,NEW}. In addition, they
also manifest that Real analyticity is recovered only for those
amplitudes which are time-reversal invariant. 

An important consequence of Hermitian analyticity is that it ensures the
equivalence between the genuine unitarity of the $S$-matrix and the
condition of `unitarity' satisfied by the $S$-matrices  derived from the
quantum group construction of refs.~\cite{QGS,MATRIX,QGToda,GEOR}. In
this construction,  the $S$-matrix is a quantum group $R$-matrix times a
Real analytic scalar factor, and the relationship between the analytic
continuations of the different amplitudes is dictated by the properties
of the $R$-matrix. As an example, using the results of Gandenberger
in~\cite{GEOR}, one can check that all the $S$-matrix amplitudes
corresponding to the affine Toda field theory associated with
$a_{2}^{(1)}$ satisfy~\foot{I thank Gustav Delius for providing a proof
that this relationship will hold for other affine Toda theories. When the
quantum parameter $q$ is a pure phase, it follows from the fact that
complex conjugation replaces $q$ by $q^{-1}$, which exchanges the two
sides in the quantum coproduct, together with the time-reversal
invariance of the $R$-matrices and the Real analyticity of the scalar
factor (see~\cite{MATRIX}).}
\be
{S_{AB}\;}_{ij}^{kl} (\theta)\> = \> \Bigl[{S_{AB}\;}_{kl}^{ij}
(-\theta^\ast)\Bigr]^\ast
\lab{ATRel}
\ee
instead of~\rf{HASgen}. This shows that the amplitudes with a
trivial matrix structure will be Real analytic, but Hermitian analyticity
will not be satisfied unless the $S$-matrix exhibits additional symmetry
properties, like parity invariance if~\rf{ATRel} holds. All this results
in the non-unitarity of these $S$-matrices reported in~\cite{TW}. 

In the same article, Tak\'acs and Watts pointed out the possibility that
some of these $S$-matrices could be conjugate to unitary matrices by
means of a rapidity-dependent change of basis of the one-particle states.
However, they found rather difficult to check it directly and proposed to
investigate instead if the two- and three-particle $S$-matrices have pure
phase eigenvalues, which is a necessary condition. Following this method,
they have singled out a number of $S$-matrix theories where such changes
of basis should exist~\cite{TW,TWnew}. Concerning this, we have
obtained a sufficient condition for the existence of a basis where
Hermitian analyticity is satisfied; we call it `weak Hermitian
analyticity', and it is summarized by eqs.~\rf{HAW} and~\rf{Split}. It
would be interesting to use weak Hermitian analyticity to characterise
those $S$-matrix theories that become unitarity in some particular basis
and, in any case, to investigate the physical meaning of such basis.

\vspace{1.5 truecm}

\noindent\centerline{\large\bf Acknowledgments} 

\vspace{0.5truecm}
I wish to thank D.~Olive for drawing my attention to the
role of Hermitian analyticity in $S$-matrix theory during the
Durham'98 TMR Conference. I would also like to thank J.~S\'anchez
Guill\'en and J.M.~S\'anchez de Santos for valuable discussions, and to
G.~Delius and G.~Watts for their clarifying comments about affine Toda
theories and the quantum group approach to factorised $S$-matrices. This
research is supported partially by CICYT (AEN96-1673), DGICYT
(PB96-0960), and the EC Commission via a TMR Grant (FMRX-CT96-0012).

\end{document}